\documentclass[a4paper, 12pt]{article}
\pdfoutput=1

\usepackage{amsmath,amssymb,amscd}
\usepackage[pdftex]{graphicx}
\usepackage{epstopdf}
\usepackage{subfigure}
\usepackage{epsfig}
\usepackage{listings}
\usepackage{caption}
\usepackage{cite}
\usepackage{dsfont}
\usepackage{slashed}
\usepackage[
    colorlinks=true,
    allcolors=blue,
    linktocpage=true
]{hyperref}

\usepackage{physics}
\usepackage{siunitx}

\newcommand{\bea}{\begin{align}}
\newcommand{\eea}{\end{align}}
\newcommand{\beq}{\begin{equation}}
\newcommand{\eeq}{\end{equation}}
\newcommand{\nbea}{\begin{align*}}
\newcommand{\neea}{\end{align*}}
\newcommand{\nbeq}{\begin{equation*}}
\newcommand{\neeq}{\end{equation*}}

\numberwithin{equation}{section}

\begin{document}

\renewcommand{\thefootnote}{\fnsymbol{footnote}}

\begin{titlepage}

\baselineskip=21pt
\rightline{KCL-PH-TH-2024-73}
\vskip 1in

\begin{center}

{\large {\bf Vacuum Metastability from Axion-Higgs Criticality}}

\vskip 0.5in

{\bf Maximilian Detering}$^{a,}$\footnote{\href{mailto:maximilian.detering@kcl.ac.uk}{maximilian.detering@kcl.ac.uk}} and {\bf Tevong~You}$^{a,}$\footnote{\href{mailto:tevong.you@kcl.ac.uk}{tevong.you@kcl.ac.uk}}

\vskip 0.3in

{\small {$^a$\it
{Theoretical Particle Physics and Cosmology Group, Department of Physics, \\ King’s College London, London, WC2R 2LS, UK}
}}

 \vskip 0.2in

 {\bf Abstract}

\end{center}


{\small

Self-organised criticality, realised through cosmological dynamics in the early universe, is an alternative paradigm for addressing the electroweak hierarchy problem. In this scenario, an unnaturally light Higgs boson is the result of dynamics driving the electroweak vacuum towards a near-critical metastable point where the Higgs mass is bounded from above by the vacuum instability scale. To lower the vacuum instability scale close to the weak scale, previous realisations of this mechanism introduced new vector-like fermions coupled to the Higgs. Here we show that an Axion-Like Particle (ALP) coupling to the Higgs is an alternative possibility for achieving criticality with another well-motivated and naturally light candidate for new physics, thus leading to an entirely different set of testable phenomenological signatures. Our Axion-Higgs criticality model predicts an ALP in the MeV to $\mathcal{O}(10)$ GeV range. The entire natural region of parameter space can be thoroughly explored by a combination of future colliders, flavour experiments, and cosmological observatories. 
 
}


\end{titlepage}
\newpage

\setcounter{footnote}{0}
\renewcommand{\thefootnote}{\arabic{footnote}}

\tableofcontents

\newpage


\section{Introduction}

The unreasonable effectiveness of the Standard Model (SM) in particle physics has put the hierarchy problem of the Higgs boson into sharp focus~\cite{Giudice:2008bi, Giudice:2013yca, Craig:2022eqo}. Prior to the discovery of the Higgs~\cite{Aad:2012tfa, Chatrchyan:2012xdj}, it was expected through naturalness arguments to be accompanied by new physics around the weak scale. This expectation was based upon the very same principles of symmetry and Effective Field Theory (EFT) responsible for our understanding of the structure of the SM and its success. It is therefore puzzling that no new physics Beyond the Standard Model (BSM) exists in the regimes of energy and precision currently being explored. This may indicate a failure in applying past EFT reasoning and symmetry principles to the Higgs sector. 

Such a failure had already been hinted at by the related cosmological constant problem~\cite{Weinberg:1988cp, Carroll:2000fy, Bousso:2007gp}. Its unnaturally small but non-zero value at such a low energy scale defies explanation in the conventional symmetry-based approach, with the most widely accepted solution being an anthropic selection in a landscape of possible values~\cite{Weinberg:1987dv}. The strong CP problem is another naturalness puzzle whose most widely accepted and theoretically appealing solution---the QCD axion~\cite{Peccei:1977hh, Weinberg:1977ma, Wilczek:1977pj, Zhitnitsky:1980tq, Dine:1981rt}---eschews symmetry in favour of cosmological dynamics. This cosmological approach to naturalness problems has been reinvigorated in the last decade~\cite{Asadi:2022njl}. Ref.~\cite{Graham:2015cka} introduced a new mechanism for relaxing the weak scale far below the EFT cut-off using an Axion-Like Particle (ALP), the relaxion, to scan the Higgs mass, with a trapping backreaction triggered by the Higgs' vacuum expectation value. Other variants (e.g.~\cite{Espinosa:2015eda,Hardy:2015laa,Batell:2015fma,Marzola:2015dia,Evans:2016htp,Hook:2016mqo,You:2017kah,Evans:2017bjs,Batell:2017kho,Ferreira:2017lnd,Tangarife:2017rgl,Davidi:2017gir, Fonseca:2017crh,Son:2018avk,Fonseca:2018xzp,Davidi:2018sii,Wang:2018ddr,Gupta:2019ueh,Fonseca:2019aux,Ibe:2019udh,Kadota:2019wyz,Fonseca:2019lmc, Domcke:2021yuz, Matsedonskyi:2023tca}) and alternative cosmological solutions have since been proposed, for example involving four-form fluxes~\cite{Giudice:2019iwl, Lee:2019efp, Moretti:2022xlc, Kaloper:2022oqv}, reheating with a large number of copies of the SM~\cite{Arkani-Hamed:2016rle},   preferentially inflating vacua with a weak-scale Higgs mass~\cite{Geller:2018xvz, Cheung:2018xnu, Chattopadhyay:2024rha} or crunching other patches of the landscape~\cite{TitoDAgnolo:2021nhd, TitoDAgnolo:2021pjo, Csaki:2024ywk}.         

More recently, models of cosmological self-organised criticality have been inspired by other physical systems that appear fine-tuned to lie perched at a near-critical point of a phase transition~\cite{Khoury:2019yoo, Khoury:2019ajl, Kartvelishvili:2020thd, Khoury:2021grg, Giudice:2021viw,Steingasser:2024hqi}. Such a self-organisation towards an equilibrium with a metastable vacuum could arise from vacuum transition dynamics in the string landscape~\cite{Khoury:2019yoo, Khoury:2019ajl, Kartvelishvili:2020thd, Khoury:2021grg} or by Self-Organised Localisation (SOL)~\cite{Giudice:2021viw} of an apeiron scalar field responsible for a quantum phase transition. A generic prediction of this mechanism applied to the Higgs mass is a near-criticality of the Higgs potential with its mass bounded from above by the vacuum instability scale. A Higgs criticality model would therefore explain the Higgs mass fine-tuning by placing us in a metastable vacuum with a weak-scale Higgs bilinear close to the edge of a phase transition. In particular, new physics coupled to the Higgs is necessary to lower this vacuum instability scale closer to the TeV range~\cite{Giudice:2021viw, Khoury:2021zao, Steingasser:2023ugv, Chauhan:2023pur, Benevedes:2024tdq}. Unlike the anthropic solution to the cosmological constant problem, a vacuum metastability bound to solve the hierarchy problem in the context of a Higgs criticality model would therefore have testable consequences. 

Candidates for naturally light new physics modifying the Higgs potential include Vector-Like (VL) fermions and shift-symmetric scalars such as an ALP (hereafter we shall use ALP and axion interchangeably). VL fermions with Yukawa couplings to the Higgs were used to lower the vacuum instability scale in Refs.~\cite{Giudice:2021viw, Benevedes:2024tdq}, while Ref.~\cite{Khoury:2021zao} considered a composite Higgs model with right-handed neutrinos. Here we study the alternative possibility of an axion coupling to the Higgs to achieve near-criticality with a lowered vacuum instability scale. Axions are generically expected in many extensions of the SM and in particular the string landscape~\cite{Arvanitaki:2009fg}, and their destabilising effect on the Higgs potential has been studied for strong first-order phase transitions \cite{harigaya2022firstorderelectroweakphase, harigaya2023alpassistedstrongfirstorder, Ke:2024lel}. As we shall see, our Axion-Higgs criticality model provides additional motivation for axion searches in the MeV to GeV range and demonstrates a wider range of potential phenomenological signatures of SOL or of other cosmological solutions to the hierarchy problem relying on the vacuum metastability bound.  

In the next Section, we begin by reviewing criticality and the vacuum metastability bound in the SM. Section~\ref{sec:axion-higgs} then derives bounds on the Higgs mass when an axion is coupled to the Higgs in our Axion-Higgs criticality model. Existing constraints and future projections on the parameter space of the model are discussed in Section~\ref{sec:constraints}, before concluding with a summary and discussion in Section~\ref{sec:conclusion}.

\section{Criticality in the Standard Model\label{sec:sm}}
We begin by reviewing criticality and the vacuum structure in the Standard Model as a warm-up for considering the vacuum metastability bound in a Higgs criticality model. The vacuum structure of a theory is crucial for defining  the critical point as the transition between different such configurations. It can be analysed from the point of view of the effective potential. For the Standard Model, we may write the effective potential as
\begin{equation}
    V_\text{eff} (H) = -\frac{1}{2} m_{\text{eff}}^2 H^2 + \frac{1}{4} \lambda_\text{eff} H^4, \label{eq:SM-effective-potential}
\end{equation}
where $m^2_{\text{eff}}$ and $\lambda_\text{eff}$ are the effective quadratic and quartic couplings including loop corrections to the effective potential, and $H$ is the neutral Higgs field (see also appendix~\ref{sec:SM-threshold-corrections}). The effective parameters can be computed in a loop expansion as
\begin{equation}
    m_{\text{eff}}^2 = \sum_{n} {m_\text{eff}^2}^{(n)} , \quad
    \lambda_\text{eff} = \sum_{n} \lambda_\text{eff}^{(n)} ,
\end{equation}
which we compute in the $\overline{\mbox{MS}}$ renormalisation scheme. We shall not adopt the common Renormalisation Group (RG)-improved effective potential where the effective potential is evolved with its RG equation from some scale $\mu_0$ up to another scale $\mu$. Instead, we run the $\overline{\mbox{MS}}$ couplings from the scale at which they are determined up to the scale $\mu = H$, effectively choosing this scale as the calculation scale. Unlike the RG-improved effective potential, which mixes terms to all orders in perturbation theory through the renormalisation group, this prescription for the effective potential ensures a consistent perturbative expansion to fixed order, see e.g.\@ Refs.~\cite{andreassen2014consistentuseeffective, andreassen2014consistentusestandard} for a detailed discussion. In particular, the fixed-order effective potential is obtained with the power counting $\lambda \sim \hbar$, which is necessary since we will be interested in studying the potential for anomalously small quartic couplings where loop corrections are comparable to the tree-level result. The effective bilinear can also be expanded perturbatively with a similar power-counting, however, the Born result only receives negligible corrections.
The leading order (LO) contribution to the effective quartic coupling is thus
\begin{equation}\label{eq:SM-effective-quartic-LO}
\begin{split}
    \lambda_\text{eff}^\text{LO} = \lambda + \frac{1}{(4\pi)^2} \left[ - 12 T^2 \left( \ln{\frac{T H^2}{\mu^2} - \frac{3}{2}} \right) + 6 W^2 \left( \ln{\frac{W H^2}{\mu^2}} - \frac{5}{6} \right) + 3 Z^2 \left( \ln{\frac{Z H^2}{\mu^2}} - \frac{5}{6} \right) \right],
\end{split}
\end{equation}
where
\begin{equation} \label{eq:TWZ}
    T \equiv \frac{1}{2} y_t^2, 
    \quad W \equiv \frac{1}{4} g_2^2, 
    \quad Z \equiv \frac{1}{4} (g_Y^2 + g_2^2),
\end{equation}
with the hypercharge gauge coupling $g_Y$ and the weak isospin gauge coupling $g_2$.

The location and structure of vacua now crucially depend on the values of the effective parameters and their evolution under the renormalisation group. The vacuum structure of the potential can be analysed by looking at the stationary points of the effective potential given by solutions to the following equation,
\begin{equation}
    0 = - m_{\text{eff}}^2 - \frac{1}{2} \dv{m_{\text{eff}}^2}{\ln{H}} + \lambda_\text{eff} H^2 + \frac{1}{4} \dv{\lambda_\text{eff}}{\ln{H}} H^2.
\end{equation}
Equivalently, this can be written via the beta functions as
\begin{equation}\label{eq:SM-extremal-condition}
    m_\text{eff}^2 + \frac{1}{2} \beta_{m^2} = \left(\lambda_\text{eff} + \frac{1}{4} \beta_\lambda \right) \mu^2,
\end{equation}
where $\mu$ is defined as the scale where the equation is satisfied.
The number of solutions to this equation tells us about the vacuum structure of the SM. Assuming $m^2_{\text{eff}}>0$ and $\lambda_\text{eff}>0$ in the IR, as is realised in the SM, the existence of a single solution to Eq.~\eqref{eq:SM-extremal-condition} corresponds to a unique non-trivial IR vacuum, the electroweak vacuum. However, Eq.~\eqref{eq:SM-extremal-condition}, because of its perturbative structure, may also admit three solutions for certain values of the parameters. These three solutions correspond to an IR minimum, an intermediate maximum, and a UV minimum. Depending on the values of the parameters, the SM may thus be in a single vacuum phase (with a unique IR vacuum) or in a two-vacua phase (with an IR vacuum and a UV vacuum, which typically have a large scale separation). The critical point between the two phases corresponds to values of the parameters such that Eq.~\eqref{eq:SM-extremal-condition} admits exactly two solutions corresponding to the UV vacuum and a saddle point of the potential in the IR.

To analyse whether the SM possesses a UV vacuum, we can focus on the effective potential at large field values. For $H^2 \gg m_H^2$, the SM effective potential is well approximated by
\begin{equation}
    V^\text{UV}_\text{eff}(H) \simeq \frac{1}{4} \lambda_\text{eff} \, H^4 \, .
\end{equation}
The existence of a sub-Planckian UV vacuum of the theory can then be determined by following the RG evolution of the effective quartic coupling $\lambda_\text{eff}$. We would conclude that the SM does not admit a sub-Planckian UV vacuum if $\lambda_\text{eff}(\mu) > 0$ up to the Planck scale, while one requirement for a UV vacuum can correspond to the condition~\footnote{This condition is not always necessary and has to be complemented by an analysis of the stabilising contributions at larger scales to be sufficient to conclude the existence of a minimum.}
\begin{equation}
    \lambda_\text{eff}(\mu_I) = 0,
\end{equation}
for some \textit{instability scale} $\mu_I$ below the Planck scale. 

\subsection{UV vacuum and RG evolution of the quartic coupling}

As we have just discussed, the existence of a UV vacuum in the Standard Model can be determined by the RG evolution of the effective quartic coupling $\lambda_\text{eff}$. At leading order, see Eq.~\eqref{eq:SM-effective-quartic-LO}, it schematically takes the form,
\begin{equation}
    \lambda_\text{eff}^\text{LO} \sim \lambda + (- y_t^4 + g_2^4 + (g_1^2+g_2)^2),
\end{equation}
where we have neglected factors of order unity for the purpose of illustration.
From the leading order approximation we therefore conclude that an intricate interplay of the Higgs quartic, Yukawa and gauge couplings is required for the realisation of a UV vacuum. In particular, the Higgs self-coupling and top quark Yukawa coupling turn out to be parametrically dominant, with the 1-loop gauge coupling contributions giving parametrically smaller corrections around the electroweak scale. The dominant effects are accordingly encapsulated in the renormalisation group trajectories of the Higgs self-coupling $\lambda$ and the top Yukawa coupling $y_t$.

Crucially, the necessary condition for the existence of a UV vacuum, that the effective quartic $\lambda_\text{eff}$ vanishes at some scale $\mu_I$, can only be achieved because of the special structure of the $\lambda$ beta function with both multiplicative and additive contributions. That is, ignoring numerical factors and sub-leading terms, its 1-loop beta function takes the form,
\begin{equation}
    \beta_\lambda \sim \lambda^2 + \lambda ( y_t^2 - \text{gauge terms}) - y_t^4 + \text{gauge terms}.
\end{equation}

The following observations can be drawn directly from this structure. For a small enough Yukawa coupling in the IR, the quartic coupling remains positive up to a large scale: either the quartic grows towards larger scales or declines slowly enough to remain positive up to a large scale. This is the stable phase of the SM and the IR vacuum is the unique vacuum of the theory. For a large enough Yukawa coupling in the IR, the quartic coupling $\lambda$, and also the effective quartic coupling $\lambda_\text{eff}$, can be driven to negative values in the UV, even at sub-Planckian scales. Eventually, $\beta_\lambda$ grows towards the UV and may become positive since the top Yukawa beta function, which at 1-loop has the qualitative form
\begin{equation}
    \beta_{y_t} \sim -y_t ( \text{gauge terms} - y_t^2),
\end{equation}
is negative as the gauge terms generally outmatch the positive contribution from the top Yukawa coupling.\footnote{This is generally the case for couplings not too different from the Standard Model values, i.e.\@ in particular if the top Yukawa coupling is not much larger than its value in the SM.}
This corresponds to the possibly metastable phase of the SM, where IR and UV vacua can coexist depending on the size of the Higgs bilinear parameter. Unless the Higgs bilinear parameter is negative and large in magnitude, the theory admits both vacua.
It is due to this feature that self-organised criticality, in the context of a landscape of values for the Higgs mass, may eventually explain the smallness of the electroweak scale, as we discuss in more detail in the next section.

\begin{figure}[t]
    \centering
    \includegraphics[width=3in]{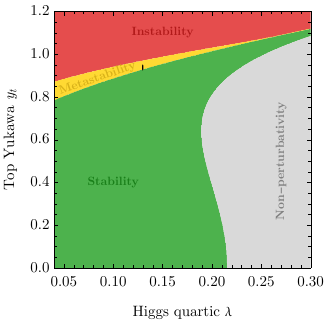}
    \includegraphics[width=3in]{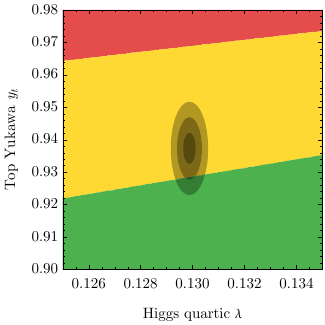}
    \caption{Standard Model phase diagram in the plane spanned by the top Yukawa coupling and Higgs quartic coupling, renormalised at the top mass scale. The measured SM values are shown with a 3-$\sigma$ ellipse on the left and with 1-, 2-, and 3-$\sigma$ contours on the right. The uncertainties are given in Eq.~\eqref{eq:couplings-at-top-scale} and include the experimental uncertainty only. The SM values for the top Yukawa and gauge couplings are given in Eq.~\eqref{eq:couplings-at-top-scale}.}
    \label{fig:SM-phase-diagram}
\end{figure}

To illustrate the possible scenarios for the Standard Model, we plot the phase diagram of the model in the plane of the Higgs quartic vs top Yukawa couplings in Fig.~\ref{fig:SM-phase-diagram}, for couplings evaluated at the top mass scale. The stable phase corresponds to the SM in which the electroweak vacuum is the unique vacuum. The meta-stable and instable phase both admit a sub-Planckian minimum in the Higgs effective potential, which we consider at leading order here. For the classification of metastability and instability we use the criterion of Ref.~\cite{buttazzo2013investigatingnearcriticalityhiggs} based on the lifetime of the electroweak vacuum that we compute numerically. The region in which the Higgs quartic coupling becomes non-perturbative ($\lambda \sim 1$) is indicated in Fig.~\ref{fig:SM-phase-diagram} as well. We ignore gravitational corrections that can be neglected when assuming a conformal coupling of the Higgs to gravity \cite{Chauhan:2023pur}.

There is still considerable uncertainty regarding the value of the instability scale. The measurements of the couplings with a finite resolution and the truncation of the perturbative expansion of both the threshold corrections and the beta functions are sources of uncertainty. Due to the exponential dependence of the instability scale as a renormalisation scale it is therefore imperative to carefully consider the uncertainty of the result. In this work we compute the couplings at next-to-leading order (NLO) with the leading corrections only, resulting in sub-percent uncertainties of the couplings at the top mass scale as the renormalisation scale, see also appendix~\ref{sec:SM-threshold-corrections}. To determine the instability scale in the SM, we use the RG equation at NNLO. We obtain for the instability scale derived from the effective quartic coupling at 1-loop,
\begin{equation}
    \mu_I = 10^{11.8^{+2.7}_{-1.4}} \, \si{GeV}.
\end{equation}
The instability scale in the literature is also sometimes approximated as the root of the Higgs quartic coupling $\lambda$, which gives us
\begin{equation}
    \mu_I = 10^{{11.1}^{+2.6}_{-1.4}} \, \si{GeV}.
\end{equation}
The uncertainty on the instability scale is therefore relatively large, with values varying over four orders of magnitude within one standard deviation.

\begin{figure}[t]
    \centering
    \includegraphics[width=5in]{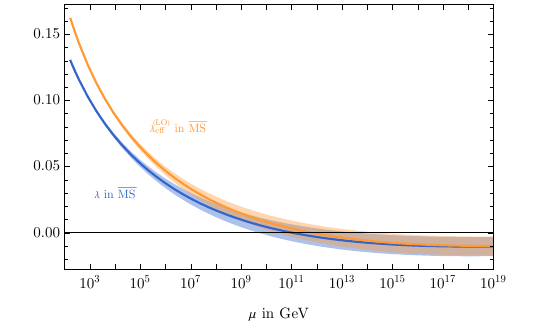}
    \caption{3-loop RG evolution of the quartic coupling $\lambda$ and effective quartic coupling $\lambda_\text{eff}$ as a function of the $\overline{\mbox{MS}}$ renormalisation scale $\mu$ with 1-$\sigma$ uncertainty bands. The used values of the couplings at the top mass scale are given in Eq.~\eqref{eq:couplings-at-top-scale}. The plot is similar to Fig.~3 in Ref.~\cite{degrassi2012higgsmassvacuum}. Note that the uncertainty in the Higgs mass measurement is significantly reduced compared to the values used in Ref.~\cite{degrassi2012higgsmassvacuum}.}
    \label{fig:quartic-coupling}
\end{figure}

The quartic coupling $\lambda$ and the effective quartic coupling $\lambda_\text{eff}$ as a function of the renormalisation scale are shown in Fig.~\ref{fig:quartic-coupling} with their respective uncertainty bands. This order of magnitude uncertainty in $\mu$ for the quartic turnover point could be reduced at future colliders~\cite{franceschini2023colliderlandscapewhich} to establish the instability scale in the SM within a factor of 2~\cite{dunsky2021snowmass2021letterinterest}, or help pinpoint any BSM modifying the running of the Higgs potential.

\subsection{Criticality and the smallness of the Higgs bilinear}

In the phase of the SM with a sufficiently small quartic coupling for a UV vacuum, which seems to be the case if one assumes no BSM, the effective potential may or may not exhibit an IR vacuum alongside the UV vacuum if understood as a function of the (effective) Higgs mass parameter $m_\text{eff}^2$. This can be illustrated as follows.

If the quadratic term is large enough and dominates up to scales where the quartic coupling already becomes negative, then there cannot be an IR vacuum but only a UV vacuum. If however the quadratic term is small, i.e.\@ the quartic term dominates already at scales where $\lambda$ is positive (below the instability scale), then the potential will have an additional IR vacuum. The two vacua are hence exponentially separated by the renormalisation group evolution of the (effective) quartic coupling. We call the value of the bilinear parameter critical if it corresponds to the transition point between the two phases.

The phase of the model can be determined from the number of solutions to the stationary condition, Eq.~\eqref{eq:SM-extremal-condition}. Using the Higgs mass anomalous dimension,
\begin{equation}
    \gamma_{m} \equiv - \dv{\ln m}{\ln \mu},
\end{equation}
and the beta function, we can solve for the Higgs mass parameter,
\begin{equation}\label{eq:SM-critical-bilinear-equation}
    m^2 \, \left( 1 + \Delta r_{m^2} + \gamma_{m}\right) =  \left( \lambda_\text{eff} + \frac{1}{4} \beta_\lambda \right) \mu^2,
\end{equation}
where we defined
\begin{equation}
    \Delta r_{m^2} = \frac{1}{m^2} \sum_{n} {m_\text{eff}^2}^{(n)},
\end{equation}
with $m^2$ the $\overline{\mbox{MS}}$ parameter and $\Delta r_{m^2}$ encompassing the radiative corrections to the mass parameter. Note that in Eq.~\eqref{eq:SM-critical-bilinear-equation} all $\overline{\mbox{MS}}$ parameters depend implicitly on the renormalisation scale $\mu$. To be precise, $m^2$ depends on $\mu$; $\gamma_m$ and $\beta_\lambda$ are functions of $\overline{\mbox{MS}}$ couplings, which in turn depend on $\mu$. And $\Delta r_{m^2}$, $\lambda_\text{eff}$ are functions of the couplings, which again implicitly depend on $\mu$. As mentioned before, we have resummed the effective potential implicitly at the scale $\mu=H$, therefore eliminating any explicit dependence on $\ln H/\mu$ in Eq.~\eqref{eq:SM-critical-bilinear-equation}.

\begin{figure}[t]
    \centering
    \includegraphics[width=4in]{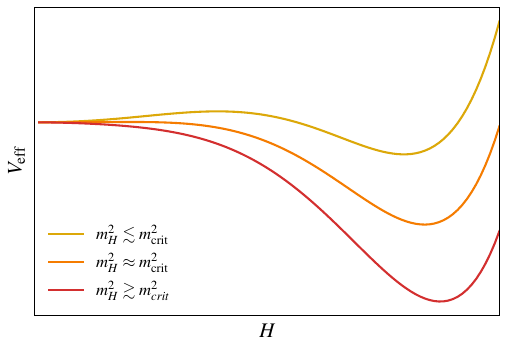}
    \caption{Sketch of the effective Higgs potential for different values of the Higgs bilinear corresponding to different phases of the theory. A subcritical value of the bilinear parameter corresponds to the metastable phase (yellow) admitting IR+UV vacuum. The supercritical bilinear results in the unstable phase (red) only admitting a UV vacuum. The critical value of the bilinear corresponds to the transition point between the two phases, where the IR vacuum turns into a saddle point of the potential (orange). The Standard Model is in the metastable phase (yellow).}
    \label{fig:1d-potential-sketch}
\end{figure}

Now, for given $\overline{\mbox{MS}}$ couplings $\lbrace g_i(\mu), y_j(\mu), \lambda(\mu) \rbrace$, we want to determine the number of solutions to Eq.~\eqref{eq:SM-critical-bilinear-equation} for a given value of $m^2$. In Ref.~\cite{degrassi2012higgsmassvacuum} it was shown that one solution exists in the far but sub-Planckian UV, corresponding to a UV vacuum of the Standard Model.\footnote{This is true if $m^2$ is sufficiently small, i.e.\@ negligible compared to the scale of the UV vacuum, which is the case in the SM.} Then, the existence of further solutions corresponding to an IR vacuum (the electroweak vacuum) and a separating maximum in the potential depends on the value of $m^2$. To this end, we solve for $m^2$,
\begin{equation}\label{eq:SM-ciritcal-bilinear-equation}
    m^2 = \mu^2 \left(\lambda_\text{eff} + \frac{1}{4} \beta_\lambda \right) \left( 1 + \Delta r_{m^2} + \gamma_{m} \right)^{-1}.
\end{equation}
We can solve this equation numerically to determine the number of sub-Planckian vacua. Solving Eq.~\eqref{eq:SM-ciritcal-bilinear-equation} at LO with running at NNLO, the critical value for the Higgs bilinear is
\begin{equation}
    m_\text{crit}^2 = \left(\SI{3.2e10}{GeV}\right)^2 .
\end{equation}
To gain further insight, we can also consider an analytic approximation for the upper bound on $m^2$ that allows the existence of an IR vacuum. To this end, note that the IR minimum can only exist below the so-called instability scale $\mu_I$, which is defined as the renormalisation scale corresponding to a vanishing effective quartic coupling,
\begin{equation}
    \lambda(\mu_I) = 0,
\end{equation}
where $\mu_I < v_\text{UV}$, $v_\text{UV}$ is the vacuum expectation value in the UV vacuum, and we have resummed the effective parameter at the scale $\mu\!=H=\!\mu_I$. Working to 1-loop and leading-log approximation, we can expand Eq.~\eqref{eq:SM-ciritcal-bilinear-equation} around the instability scale,
\begin{equation}\label{eq:SM-critical-bilinear-one-loop}
    m^2 = \left.\beta_\lambda^{(1)}\right\rvert_{\mu_I} \left( \ln\frac{\mu}{\mu_I}  + \frac{1}{4} \right) \mu^2 .
\end{equation}
Note that $\Delta r_{m^2}$ and $\gamma_m$ first contribute at two-loop order. Eq.~\eqref{eq:SM-critical-bilinear-one-loop} takes the simple form,
\begin{equation}\label{eq:xexpx}
    \rho = \xi \; \mathrm{e}^{\xi},
\end{equation}
with
\begin{align}
    \rho &\equiv \frac{2 m^2}{\mu_I^2} \frac{\sqrt{\mathrm{e}}}{\beta_\lambda^{(1)}\vert_{\mu_I}}, \\
    \xi &\equiv \ln \frac{\mu^2 \sqrt{\mathrm{e}}}{\mu_I^2}.
\end{align}
The solutions are given by Lambert's $W$-functions. In particular, since $\beta_\lambda < 0$ for $\mu\sim\mu_I$ and $\rho < 0$, the solutions are given by $W_{-1}$. The minimal value of $\rho$, and thus the maximal value of $m^2$ for which a solution exists is $\rho = -\mathrm{e}^{-1}$. Thus,
\begin{equation}
    m_\text{crit}^2 = - \frac{1}{2} \left.\beta_\lambda^{(1)}\right\rvert_{\mu_I} \mathrm{e}^{-3/2} \; \mu_I^2 \, .
\end{equation}
This critical parameter corresponds to a saddle point in the potential, i.e.\@ the transition point between the UV vacuum phase and IR+UV vacuum phase of the model. For values of $m_{\text{eff}}^2 > m_\text{crit}^2$, the IR vacuum disappears. For the potential given in Eq.~\eqref{eq:SM-effective-potential}, the requirement of a saddle point corresponds to a vanishing second derivative or equivalently vanishing mass for the fluctuations around the saddle point. A sketch of the potential for the different cases is shown in Fig.~\ref{fig:1d-potential-sketch}.

Criticality in the Standard Model can therefore reduce the hierarchy problem in the EFT paradigm~\footnote{At least if we take the scale of new physics, $\Lambda$, to be much larger then the instability scale. As mentioned, we assume $\Lambda$ to be of the size of the Planck scale.}, $m_H^2 \ll \Lambda^2$, to a smaller hierarchy $m^2 \ll m^2_{\text{crit}} \lesssim \mu_I^2 \ll \Lambda^2$ since the instability scale through logarithmic RG running can naturally be exponentially far below from the UV scale $\Lambda$.\footnote{See Ref.~\cite{khoury2021gaugehierarchyelectroweak} for a detailed discussion.}\footnote{Care should be taken when interpreting the involved scales as physical. The renormalisation scale, and hence in particular the instability scale, are unphysical scales. However, comparing the $\overline{\text{MS}}$ renormalisation scale $\mu$ to other $\overline{\text{MS}}$ parameters such as the effective bilinear parameter $m^2$ results in intrinsically consistent statements. Physical interpretations of these scales should only be taken in the way they are related to physical observables, such as experimentally measured cross sections.} However, the instability scale in the SM, though hierarchically smaller than the Planck scale, still turns out to be much larger than the electroweak scale with $\mu_I \approx \SI{e11}{\GeV}$. Previous work has addressed this issue by lowering the instability scale through new physics contributions to the beta function of the Higgs quartic from vector-like fermions and higher dimensional operators. In the next section we will explore an alternative explanation of the remaining hierarchy based on an axion-like particle coupled to the Higgs.

\section{Criticality in the Axion-Higgs model\label{sec:axion-higgs}}

We consider an extension of the Standard Model by an axion-like particle. From a bottom-up point of view, this is motivated by being one of the candidates for new BSM fields that can be naturally light. From a top-down perspective, axions are generically expected in string theory~\cite{Arvanitaki:2009fg}. If the vacuum metastability bound is applied in the context of a string landscape, as for example in the self-organised critical multiverse~\cite{Khoury:2019ajl, Kartvelishvili:2020thd, Khoury:2021grg}, the Axion-Higgs framework may thus provide a more realistic example of a Higgs criticality model.    

\subsection{The Axion-Higgs model}
The scalar potential with the leading-order coupling between the Higgs $H$ and the axion-like particle $S$ is~\cite{harigaya2022firstorderelectroweakphase, harigaya2023alpassistedstrongfirstorder}
\begin{equation}
    V(H,S) = - \frac{1}{2} m_H^2 H^2 + \frac{1}{4} \lambda H^4 + m_S^2 f^2 \left(1 - \cos\left(\frac{S}{f}\right) \right) - \frac{1}{2} A f (H^2 - v^2) \cos\left(\frac{S}{f} - \delta\right),
\end{equation}
where $v$ is the vacuum expectation value of the Higgs field.\footnote{A scalar coupling to the Higgs has also been considered for the purpose of stabilising the electroweak vacuum with threshold effects from the heavy scalar, see e.g. Ref.~\cite{Elias-Miro:2012eoi}. In our case, we are interested in the opposite scenario of a light scalar to destabilise the Higgs potential.}
The phase difference $\delta$ between the potential of $S$ and the interaction term with the Higgs can arise from CP-violation in the UV completion of the model and is crucial for sufficiently destabilising the Axion-Higgs potential. Possible UV completions giving rise to the CP violating Axion-Higgs coupling have been discussed in the literature, see e.g.\@ Refs.~\cite{harigaya2023alpassistedstrongfirstorder, jeong2019axionicelectroweakbaryogenesis} for two such possible UV completions. We have chosen the vacuum expectation value of the additional scalar to vanish by a shift in $S$.
For large decay constants $f$, the scalar tree-level potential simplifies,
\begin{equation}
    V(H,S) = -\frac{1}{2} m_H^2 H^2 + \frac{1}{4} \lambda  H^4 + \frac{1}{2} m_S^2 S^2 - \frac{1}{2} A' S H^2 + \frac{1}{2} A' v^2 S ,
    \label{eq:tree-level-potential}
\end{equation}
with $A' \equiv A \sin{\delta}$ and a redefinition of the Higgs bilinear. By a shift in $S$, we choose the vacuum expectation value of $S$ to vanish, with $H=v$ the minimum (to all orders) of the effective potential. Notably, the linear coupling of the new scalar with the Higgs destabilises this minimum. This however does not affect the global stability of the potential for large field values, which is restored by higher order terms in $S/f$.

The leading-order potential with the modified power counting from Sec.~\ref{sec:sm} includes the gauge boson and Yukawa contributions. Neglecting subdominant contributions, i.e.\@ all Yukawa coupling contributions except for the top, the corrections are the same as in Eq.~\eqref{eq:SM-effective-quartic-LO}. There are no corrections from the scalar $S$ at leading order.

\subsection{Axion-Higgs criticality}

Criticality can also be achieved in a theory with a two-particle scalar sector. The SM example with a single scalar field was particularly simple because of the gauged $\mathrm{SU}(2)_L$ and the restriction to relevant and marginal operators. 

Generally, the critical values of the parameters have been identified by analysing the vacuum structure of the theory. In the case of the Standard Model, its phase was determined by the size of the negative bilinear relative to the so-called instability scale. In the Axion-Higgs model, this instability scale is still present and remains unchanged since by dimensional analysis the effective quartic coupling $\lambda_\text{eff}$ does not receive any corrections from the axion coupling, and the running of renormalisable couplings is not influenced by super-renormalisable couplings.

\begin{figure}[t]
    \centering
    \includegraphics[width=6in]{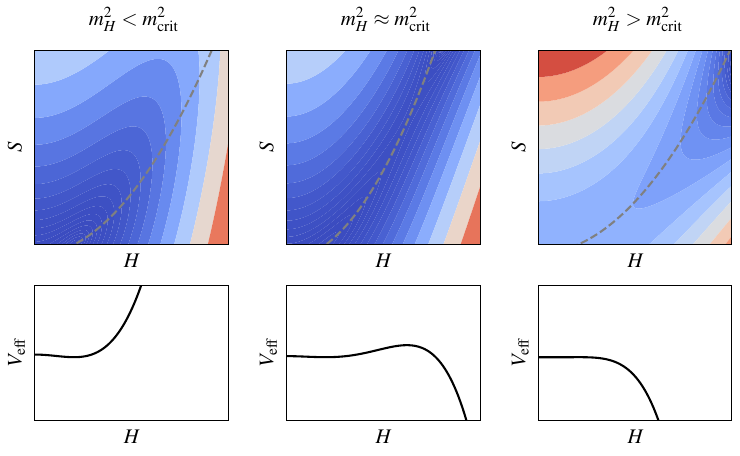}
    \caption{Effective scalar potential in the Axion-Higgs model. The first row shows the leading-order effective potential for sub-, super- and near-critical values of the Higgs bilinear. The second row shows the effective potential along the dashed line (cf.\@ first row plots) corresponding to the flat direction $\pdv{V_\text{eff}}{S}=0$.}
    \label{fig:2d-potential-sketch}
\end{figure}

However, the negative trilinear term in the potential destabilises the potential in the IR for non-vanishing values of $S$. Depending on the size of the Higgs bilinear relative to the Higgs quartic and the Axion-Higgs trilinear coupling, the model can again admit either only a UV vacuum for super-critical values of the Higgs bilinear or IR \textit{and} UV vacuum for sub-critical values. In this model, the critical point corresponding to the transition between both phases can be lowered with respect to the SM.
The effective potential in the IR+UV vacuum phase, close to the critical point, and for the UV vacuum phase of the model only is shown in Fig.~\ref{fig:2d-potential-sketch}.
The dotted line shows the curve along which $\pdv{V_\text{eff}}{S} = 0$. Since $\pdv{V_\text{eff}}{H}$, whose sign is essentially given by the sign of the quartic coupling, is negative in the UV below the UV vacuum scale, an IR vacuum can only be realised if this derivative becomes positive in the IR. Requiring an IR vacuum therefore translates to a bound on the Higgs bilinear if the remaining parameters are kept fixed. It is essentially bounded by the scale at which the derivative $\pdv{V_\text{eff}}{H}$ becomes negative along the flat $S$-direction, i.e.\@ the dotted trajectory in Fig.~\ref{fig:2d-potential-sketch}. This is what we call the instability scale in the Axion-Higgs model. The potential along this flat direction is shown in the lower part of Fig.~\ref{fig:2d-potential-sketch}.
For the Axion-Higgs model, we write the LO effective potential directly as,
\begin{multline}\label{eq:Axion-Higgs-LO-effective-potential}
    V^{\text{LO}}_\text{eff}(H,S) = 
    - \frac{1}{2} m_{H}^2 H^2 + \frac{1}{2} m_{S}^2 S^2 -\frac{1}{2} A' S H^2 + \frac{1}{4} \lambda H^4 \\
    + \frac{H^4}{4} \frac{1}{(4\pi)^2} \left[  -12 T^2  \left( \ln \frac{T H^2}{\mu^2} - \frac{3}{2} \right) 
    + 3 Z^2 \left(\ln{\frac{ZH^2}{\mu^2}} - \frac{5}{6} \right) + 6 W^2 \left(\ln{\frac{W H^2}{\mu^2}} - \frac{5}{6} \right) \right],
\end{multline}
with $T,Z,W$ from Eq.~\eqref{eq:TWZ}.
The conditions for a stationary point in the 2-scalar effective potential are given by
\begin{equation}
\begin{split}
    m_H^2 &= \lambda_\text{eff} H^2 + \frac{1}{4} \dv{\lambda_\text{eff}}{\ln H} H^2 - A^{\prime} S \\
    m_S^2 S &= \frac{1}{2} A^{\prime} (H^2 - v^2)
\end{split}
\end{equation}
We can combine these by eliminating $S$,
\begin{equation}
    m_H^2 = \left( \lambda_\text{eff}(0) - \frac{1}{2} \frac{A'^2}{m_S^2}  + \frac{1}{4} \beta_\lambda^{(1)} \right) \mu^2 + \frac{1}{2} \frac{A'^2}{m_S^2} v^2,
\end{equation}
where we have evaluated the effective potential at $\mu=H$. Note that the contribution to $\beta_\lambda^{(1)}$ corresponding to Eq.~\eqref{eq:Axion-Higgs-LO-effective-potential} are only due to the electroweak gauge couplings and the top Yukawa coupling. Analogous to the SM, we can expand this equation around the instability scale, which we now define as the renormalisation scale of vanishing reduced effective quartic coupling $\Tilde{\lambda}_\text{eff}$,
\begin{equation}
    \Tilde{\lambda}_\text{eff} \equiv \lambda_\text{eff} - \frac{1}{2} \frac{A'^2}{m_S^2}.
\end{equation}
Then, in the leading-log approximation,
\begin{equation}
    m_H^2 = \beta_\lambda^{(1)} \mu^2 \left( \ln \frac{\mu}{\mu_I} + \frac{1}{4} \right) + \frac{1}{2} \frac{A'^2}{m_S^2} v^2,
\end{equation}
or, equivalently,
\begin{equation}
    \Tilde{\rho} = \Tilde{\xi} \; \mathrm{e}^{\Tilde{\xi}},
\end{equation}
where
\begin{align}
    \Tilde{\rho} &= \frac{2\sqrt{\mathrm{e}}}{\beta_\lambda^{(1)} \mu_I^2} \left( m_H^2 - \frac{1}{2} \frac{A'^2 v^2}{m_S^2} \right)  \\
    \Tilde{\xi} &= \ln{\frac{\mu^2\sqrt{\mathrm{e}}}{\mu_I^2}}
\end{align}
The critical value for the bilinear parameter then follows as previously,
\begin{align}
    m^2_{\text{crit}} = - \frac{1}{2} \left.\beta_{\lambda}\right\rvert_{\mu_I} \mathrm{e}^{-3/2} \mu_I^2 + \frac{1}{2} \frac{A'^2 v^2}{m_S^2}.
\end{align}
The critical value for the bilinear is therefore typically one order of magnitude smaller than the instability scale.

The hierarchy between the scale of new physics---which can for example be taken to be the Planck scale $M_\text{Pl}$ or the grand unification scale $\Lambda_\text{GUT}$---and the electroweak scale is thus explained in the Axion-Higgs model (assuming an underlying theory of self-organised criticality \cite{Khoury:2019yoo, Khoury:2019ajl, Kartvelishvili:2020thd, Khoury:2021grg, Giudice:2021viw,Steingasser:2024hqi}) by the logarithmic running of the effective quartic at the UV scale $\Lambda$ and its value ${A'^2}/{2m_S^2}$ in the IR. For a sizeable value of this ratio, that is if it is smaller than but comparable to the effective quartic coupling measured at the electroweak scale, the bound on the Higgs bilinear can be very restrictive. In fact, the bound is so restrictive that not all values for $A'$ and $m_S^2$ would lead to a vacuum at the electroweak scale.

\begin{figure}[t]
    \centering
    \includegraphics[width=6in]{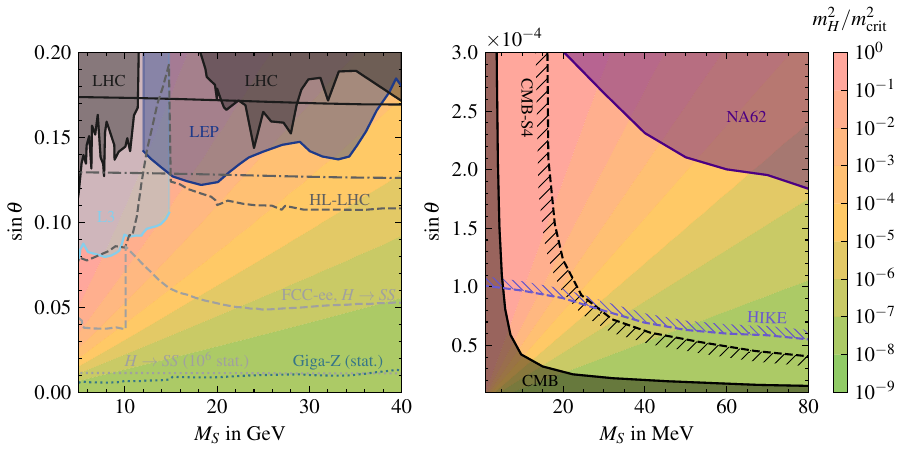}
    \caption{Contours of the ratio of the observed Higgs bilinear, $m_H^2$, to the critical value of the Higgs bilinear, $m^2_\text{crit}$, in the plane of the scalar mass $M_S$ and the sine of the mixing angle $\sin{\theta}$. Here, the Higgs mass parameter is calculated for the given parameter point in the ALP parameter space. Current constraints are shown by shaded regions and solid lines. Future projections are denoted by dashed lines, and dotted lines for statistical limits only. Current upper limits from the global fits on the Higgs exotic branching ratio at 16\% \cite{atlascollaboration2021combinedmeasurementshiggs} are shown by a solid line (black) and expected bounds at 4\% \cite{deBlas:2019rxi} by the lower dash-dotted line.}
    \label{fig:critical-vev}
\end{figure}

To analyse this scenario, we compute the critical Higgs vacuum expectation value, ${m}_\text{crit}^2$, for different axion parameters. To this end, we translate the observables, in particular $\lbrace M_H, M_S, \sin{\theta} \rbrace$, where $\theta$ is the $H-S$ mixing angle, to the renormalised $\overline{\mbox{MS}}$ parameters $\lbrace A', \lambda, m_H^2, m_S^2\rbrace$ at one loop. The details are described in Appendix~\ref{sec:threshold-corrections}.
The results are shown in Fig.~\ref{fig:critical-vev} for the parameter space with masses in the $\si{MeV}$ to $\si{GeV}$ range in the left and right plots respectively vs the mixing angle. The colour-coded contours denote the ratio $m_H^2 / m_\text{crit}^2$ which can be taken as a measure of fine-tuning for the observed Higgs mass in the parameter space of our Axion-Higgs criticality model. The relatively natural region, allowing for a little hierarchy up to two orders of magnitude separation between $m_H^2$ and the critical bilinear, lies in the range of ALP masses between an MeV and 10 GeV corresponding to mixing angles $\sin\theta$ from $10^{-4}$ to $10^{-1}$. Heavier axions or smaller mixings are also possible if one accepts a larger degree of fine-tuning. In the decoupled limit this ratio tends towards the SM result as expected.   

\section{Phenomenological signals and constraints\label{sec:constraints}}

An axion-like particle mixing with the Higgs boson has several phenomenological implications and signatures that can be probed by particle physics experiments and cosmology. Part of the parameter space is already constrained and future proposed experiments would be able to rule out or discover a signal in the entire natural region of parameter space. This is shown by the current constraints in shaded regions and solid lines, and projected constraints in dashed and dotted lines respectively in Fig.~\ref{fig:critical-vev}, which we discuss in this Section. A review of various constraints on the Axion-Higgs model has previously been conducted in Ref.~\cite{harigaya2023alpassistedstrongfirstorder}, while the general case of a singlet scalar mixing with the Higgs has been studied, for example, in Ref.~\cite{antel2023feeblyinteractingparticles}. As for theoretical constraints, we have explicitly checked that perturbative unitarity bounds are irrelevant in the considered region of parameter space and that the metastable vacuum is sufficiently long-lived.

\subsection{Exotic Higgs decays}
If the axion-like particle is light enough, $2 M_S < M_h$, then the Higgs can decay via $h \to S S$. As the Higgs boson has a very small decay width $\Gamma_h \approx \SI{4}{MeV}$, even tiny new physics corrections can have a detectable impact. Such Higgs exotic decays have extensively been searched for at the LHC, see for example Ref.~\cite{carena2023probingelectroweakphase} for a recent review. 
Current global fits constrain the Higgs exotic decay branching ratio to $\operatorname{BR}(h\to SS) \leq 0.16$ at 95\% C.L.\@ \cite{atlascollaboration2021combinedmeasurementshiggs}, with an expected future upper bound at 4\% \cite{deBlas:2019rxi}. The $h \to SS$ decay can lead to various final states depending on the subsequent decay channels of the light scalar $S$. In the minimal extension with just the singlet scalar, the scalar inherits the Higgs-like hierarchy in the branching fractions to fermions due to the mixing. Therefore, the final state is dominated by $b$-quark, jet, and $\tau$ final states depending on the mass of the light scalar. If the BSM extension contains further particles that the light scalar can decay into or is stable on collider timescales, the signature would be an invisible Higgs decay.

The effective coupling for the $hSS$ vertex after ALP-Higgs mixing relevant for the decay at leading order is given by
\begin{equation}
    g_{hSS} = 6 \lambda v \sin^2\theta \cos\theta -  A (2 \sin\theta - 3 \sin^3\theta)
\end{equation}
Thus, the exotic decay rate is at tree-level
\begin{equation}\label{eq:higgs-exotic-branching-ratio}
    \Gamma_{h\to SS} = \frac{g_{hSS}^2}{16\pi M_h^3} \lambda^{1/2} (M_h^2, M_S^2, M_S^2),
\end{equation}
where $\lambda(a,b,c)=(a-b-c)^2-4bc$ is the Källén function.
Note that this effective coupling vanishes in the limit of small mixing angles.\footnote{Recall that we are working in the limit $A'/f \to 0$. Non-vanishing contributions for negligible mixing angle are suppressed by the ALP decay constant $f$.}

Bounds on the branching ratio $\operatorname{BR}(h \to SS)$ can be derived from the total decay rate into the SM final states if we assume a Higgs-like hierarchy caused by the mixing. Current bounds from the LHC and future projections from HL-LHC as well as a future Higgs factory are obtained in Ref.~\cite{carena2023probingelectroweakphase}. Here we take the most constraining channel as a bound on the branching ratio and translate it into a bound on the mixing angle using Eq.~\eqref{eq:higgs-exotic-branching-ratio}. The constraints and projected sensitivities for Higgs exotic decays are shown in black and grey colours in the left panel of Fig.~\ref{fig:critical-vev}. The LHC excluded regions are shaded in black on the left plot of Fig.~\ref{fig:critical-vev}, and the projected reach for $H \to SS$ of HL-LHC, FCC-ee, and a Higgs factory assuming only a statistical limit with $10^6$ Higgs bosons is shown with grey dashed and dotted lines respectively. Indirect constraints from the Higgs exotic branching ratio are indicated by the black solid line for LHC and grey dash-dotted line for HL-LHC. We see that future colliders, sensitive to the parameter space above the exclusion lines, will be crucial for probing the entire natural region of parameter for GeV-scale ALPs in our Axion-Higgs criticality model.

\subsection{Scalar direct production}
The mixing of the interaction eigenstates to the mass eigenstates induces a coupling between the SM gauge bosons and the ALP mass eigenstate. In particular, a coupling for the vertex $SZZ$ is generated, allowing for direct production of the scalar at Z factories.

Multiple searches for direct production were performed at LEP. The L3 collaboration search for the Higgs boson~\cite{acciarri1996searchneutralhiggs} can be reinterpreted as a constraint on the ALP. In particular, a decay-independent analysis was performed for masses $0 < M_S < \SI{15}{GeV}$ via the $H \nu \Bar{\nu}$ channel. The resulting constraints from this Higgs search has been reinterpreted for the ALP in Ref.~\cite{acciarri1996searchneutralhiggs} and is denoted by the blue-shaded region in Fig.~\ref{fig:critical-vev}. 

For heavier masses, the combined searches of all LEP collaborations for the Higgs boson \cite{alephcollaboration2003searchstandardmodel} assuming decays of the scalar with SM-like couplings, and exclusive decays into $b\Bar{b}$ or $\tau\Bar{\tau}$ final states, lead to exclusion limits on the mixing angle up to masses below $\SI{110}{GeV}$. The results from Ref.~\cite{alephcollaboration2003searchstandardmodel} directly translate to bounds on $\sin^2{\theta}$. The LEP limits are shaded in dark blue in Fig.~\ref{fig:critical-vev}. Furthermore, the OPAL collaboration has performed decay-independent searches for new scalar bosons via the Bjorken process $e e \to S Z$ \cite{opalcollaboration2003decaymodeindependentsearches}. The constraints are weaker and so not shown here.

Future Z-pole factories will also be able to tightly constrain a singlet scalar mixing with the Higgs. In Fig.~\ref{fig:critical-vev} we show projections for a Giga-Z machine from a naive statistical rescaling of the constraints from LEP experiments, following Ref.~\cite{fuchs2021collidersearchesscalar}. We see that the entire GeV parameter space could be explored already with Giga-Z statistics, a conclusion that would be strengthened at a Tera-Z factory.  

\subsection{Rare meson decay}
The Axion-Higgs mixing also leads to extra decay channels of mesons, which can be searched for at $B$- and $K$-factories, leading to constraints for lighter masses in the MeV to few GeV range. For comprehensive and recent reviews, see Refs.~\cite{beacham2019physicscolliderscern, antel2023feeblyinteractingparticles}.
The following decay channels are used to search for exotic decays to singlet scalars:
\begin{align*}
    &B \; \to K \; S \; \to \; K l^{+} l^{-} \, , \\
    &K \; \to \pi \; S \; \to \pi \; l^{+} l^{-} \, , \\
    &K^+ \to \pi^+ + \text{invisible} \, .
\end{align*}
See in particular Ref.~\cite{antel2023feeblyinteractingparticles} for a comprehensive compilation of recent constraints and projections. We use the experimental constraints from NA62 Run 1 \cite{na62collaboration2021searchfeeblyinteracting} and show future projections from HIKE~\cite{cortinagil2022hikehighintensity} as a representative benchmark for the potential of future flavour experiments. 

The exotic decay rate of kaon decay into a pion and singlet scalar is given by~\cite{winkler2019decaydetectionlight}
\begin{equation}
    \Gamma (K^{+} \to \pi^{+} S) \simeq \frac{\abs{g_{Sds}}^2}{16\pi M_{K^+}^3} \left\lvert \bra{\pi} \Bar{d}_L s_R \ket{K} \right\rvert^2 \lambda^{1/2}(M_{K^+}^2,M_{\pi^+}^2,M_S^2).
\end{equation}
The matrix element can be approximated as \cite{kamenik2012fcncportalsdark}
\begin{equation}
    \left| \bra{\pi} \Bar{d}_L s_R \ket{K} \right| \simeq \frac{1}{2} \frac{M_{K^+}^2 - M_{\pi^+}^2}{M_s - M_d}.
\end{equation}
The scalar effective flavour-violating coupling is obtained by integrating out the $W$-top-loop \cite{batell2011multileptonsignatureshidden}
\begin{equation}
    g_{Sds} = \frac{3 (\sqrt{2}G_F)^{3/2}}{16\pi^2} \sin\theta M_s M_t^2 V_{td}^{*}V_{ts},
\end{equation}
where $V$ is the CKM matrix.
The decay rate for the rare $B$-meson decay is given by \cite{winkler2019decaydetectionlight}
\begin{equation}
    \Gamma (B^{+} \to K^{+} S) \simeq \frac{\abs{g_{Ssb}}^2}{16\pi M_{B^{+}}^3} \left\lvert \bra{K} \Bar{s}_L b_R \ket{B} \right\rvert^2 \lambda^{1/2}(M_{B^{+}}^2,M_{K^{+}}^2,M_S^2),
\end{equation}
where the matrix element can be approximated as \cite{ball2005newresultspion}
\begin{equation}
    \left|\bra{K}\Bar{s}_L b_R \ket{B}\right| \simeq \frac{1}{2} \frac{M_{B^{+}}^2-M_{K^{+}}^2}{M_b-M_s} f_K(M_S^2),
\end{equation}
with the form factor \cite{ball2005newresultspion}
\begin{equation}
    f_K(q^2) = \frac{0.33}{1-q^2/\SI{37.5}{\square\GeV}}.
\end{equation}
The effective flavour violating coupling is \cite{batell2011multileptonsignatureshidden}
\begin{equation}
    g_{Ssb} = \frac{3 (\sqrt{2}G_F)^{3/2}}{16\pi^2} \sin\theta M_b  M_t^2 V_{ts}^{*}V_{tb}.
\end{equation}

The current and future flavour constraints from rare meson decays are denoted respectively by the purple-shaded region and the dashed line (excluding everything above the line) on the right plot of Fig.~\ref{fig:critical-vev}.

\subsection{CMB bounds on \texorpdfstring{$N_\text{eff}$}{Neff}}
Complementary to collider constraints are cosmological constraints on the effective number of relativistic degrees of freedom $N_\text{eff}$ through the effect of axion-like particles on neutrino decoupling for low enough ALP masses. The available axion sector depletes entropy from the SM bath, therefore reducing the effective number of relativistic species, which is tightly constrained by CMB observations, see e.g. Ref.~\cite{navas2024reviewparticlephysics} for details.

The effect of axions on $N_\text{eff}$ depends decisively on the production, self-interactions and lifetime. In our model, self-interactions of the axion, which typically weaken the constraints from $N_\text{eff}$, are absent in the low energy theory. Short-lived ALPs with $\mathcal{O}(\si{MeV})$ masses can be thermalised and modify the neutrino decoupling. The effect of long-lived ALPs can be even larger due to non-thermal imprints from ALP decays at later times on the CMB \cite{ibe2022cosmologicalconstraintsdark}. The production of axions depends also on the reheating temperature. Here, we take the most conservative constraints, which come from low reheating temperatures but not too low to be in conflict with Big Bang Nucleosynthesis.

We adopt the constraints derived in Ref.~\cite{ibe2022cosmologicalconstraintsdark} (see Fig.~17 therein in particular) for a scalar mixing with the Higgs boson from Planck 2018 data \cite{planckcollaboration2020planck2018results} and projected constraints for the proposed CMB-S4 observatory~\cite{abazajian2016cmbs4sciencebook}.
The results are shown as the black-shaded region for current Cosmic Microwave Background (CMB) constraints and as black dashed lines potentially excluding the parameter space below the line for CMB-S4 on the right panel of Fig.~\ref{fig:critical-vev}. We see the complementarity of future cosmological observations with future flavour experiments such as HIKE in probing the entire region of natural parameter space for an MeV-scale Axion-Higgs criticality model.

\section{Conclusion\label{sec:conclusion}}

The vacuum metastability bound is an upper limit on the Higgs mass set by the vacuum instability scale. In the context of cosmological self-organised criticality, where the Higgs mass varies over a wide range of values up to the EFT cut-off, for example in a string landscape or while scanned by another scalar, the vacuum metastability bound could explain the hierarchy problem in the absence of a symmetry-based solution of new physics at the weak scale. Should such a cosmological scenario be the answer, it predicts a different signature of relatively light and accessible new physics coupled to the Higgs that is necessary to lower the vacuum instability scale. 

We have investigated the role of an axion coupled to the Higgs in setting a novel vacuum metastability bound. This axion-like particle is not to be confused with the relaxion of Ref.~\cite{Graham:2015cka} or the apeiron of Ref.~\cite{Giudice:2021viw} that scan the Higgs mass; in our Axion-Higgs model the axion is responsible for destabilising the Higgs potential, and we remain agnostic regarding the underlying self-organised critical mechanism in which a metastable electroweak vacuum is preferentially selected (see e.g. Refs.~\cite{Khoury:2019yoo, Khoury:2019ajl, Kartvelishvili:2020thd, Khoury:2021grg, Giudice:2021viw} for some concrete mechanisms). The effect of the specific dynamical selection model factorises from the sector responsible for lowering the instability scale, though some model-dependent considerations may arise in how the parameters are being scanned. We assume here for simplicity that only the Higgs mass parameter varies significantly while the other parameters can be essentially fixed. This expectation is borne out in specific scenarios such as SOL~\cite{Giudice:2021viw}, but the vacuum metastability bound could apply more generally in a variety of different contexts for self-organised criticality.

We find that an axion in the MeV to 20 GeV range with a mixing angle of $\sin\theta$ from $10^{-4}$ to $10^{-1}$ can lower the vacuum instability scale to within two orders of magnitude of the Higgs bilinear. The heavier GeV region of parameter space can be probed by exotic Higgs decays and Z boson measurements, while the lighter MeV range is covered by cosmic microwave background constraints on the light effective number of relativistic degrees of freedom and rare meson decays in flavour experiments. Remarkably, the entire natural region of parameter space could be probed by a combination of future colliders, next-generation flavour experiments and proposed cosmological observatories that would either discover or rule out the possibility of Axion-Higgs criticality.  

This conclusion may hold more generally. Destabilising further the Higgs potential to lower the vacuum instability scale necessarily requires new physics coupled sufficiently strongly to the Higgs, implying a finite and accessible parameter space. Vector-like fermions and shift-symmetric scalars may be the only naturally light candidates for destabilising new physics close to the weak scale, assuming the Higgs boson to be the only unnaturally fine-tuned light scalar. Given the limited number of possibilities and parameter space, it would be interesting to investigate the extent to which {\it any} vacuum metastability bound as an explanation of the hierarchy problem can be systematically explored by next-generation measurements and observations. We leave this tantalising prospect to future work.

\section*{Acknowledgments}

We thank Keisuke Harigaya and Isaac Wang for helpful correspondence, and Thomas Steingasser for useful comments on an earlier version of the paper. MD is supported by a faculty studentship. TY is supported by United Kingdom Science and Technologies Facilities Council (STFC) grant ST/X000753/1.

\appendix

\section{Electroweak threshold corrections}\label{sec:threshold-corrections}
\subsection{Standard Model}\label{sec:SM-threshold-corrections}
We write the Standard Model potential for the Higgs doublet $\mathbf{H}$ in unitary gauge as
\begin{equation}
    V = - m_H^2 \abs{\mathbf{H}}^2 + \lambda \abs{\mathbf{H}}^4, \quad \mathbf{H} = \frac{1}{\sqrt{2}} \begin{pmatrix}
        0 \\ H
    \end{pmatrix}, \quad H = v + h
\end{equation}
such that, ignoring negligible width effects, the Higgs pole mass $M_h$ is the solution to the pole equation,
\begin{equation}
    M_h^2 = - m_H^2 + 3 \lambda v^2 + \Pi_{HH}(M_h^2),
\end{equation}
where $\Pi_{HH}(p^2)$ is the Higgs self-energy with external momentum $p$. Equivalently,
\begin{equation}\label{eq:pole-mass-equation}
    M_h^2 = \left( -m_H^2 + 3\lambda v^2 + \Pi_{HH}(0) \right) + \left( \Pi_{HH}(M_h^2) - \Pi_{HH}(0) \right) = M_{h,V}^2 + \Delta \Pi_{HH}(M_h^2),
\end{equation}
where we can express the first term via the effective potential,
\begin{equation}
    M_{h,V}^2 \equiv \left.\frac{\partial^2 V_\text{eff}}{\partial H^2}\right\rvert_{H=v},
\end{equation}
and we defined
\begin{equation}
    \Delta \Pi_{HH}(M_h^2) \equiv \Pi_{HH}(M_h^2) - \Pi_{HH}(0).
\end{equation}
To compute Eq.~\eqref{eq:pole-mass-equation} at one-loop, we can compute the first term from the 1-loop effective potential given by Eq.~\eqref{eq:SM-effective-potential}. In the second term at one-loop, $\Delta \Pi_{HH}^{(1)}(M_h^2)$, we can substitute the Born results $M_h^2 = 2 \lambda v^2$, so that
\begin{equation}
    \Delta \Pi^{(1)}_{HH} (M_h^2) \approx \Pi^{(1)}_{HH}(2\lambda v^2) - \Pi^{(1)}_{HH} (0).
\end{equation}
Therefore the second term at one-loop will be suppressed by another factor of $\lambda$ and subdominant. We will thus neglect this term.
The perturbative expression for $M_h^2$ in terms of $\lambda$ can be inverted to obtain the quartic coupling in terms of $M_h^2$ and other couplings. Further, to express $\lambda$ purely in terms of physical quantities (the Fermi constant and the pole masses $M_t, M_W, M_Z$), we need the relations between these physical parameters and their related $\overline{\mbox{MS}}$ couplings. At the level of accuracy we are working at, only the relations between $y_t$ and $M_t$, as well as $v$ and $G_\mu$ are required. They are, using $G_F = \frac{1}{\sqrt{2}v^2}$, at one loop:
\begin{align}\label{eq:SM-threshold-corrections}
    \lambda(\mu) &= \frac{G_F M_h^2}{\sqrt{2}} - \frac{2 G_F^2}{(4\pi)^2} \left[ 6 M_W^4 \left(L_W + \frac{2}{3}\right) + 3 M_Z^4 \left(L_Z + \frac{2}{3}\right) - 12 M_t^4 L_T \right], \\
    m_H^2 (\mu) &= \frac{M_h^2}{2} - \frac{\sqrt{2} G_F}{(4 \pi)^2} \left[ 6 M_W^4 + 3 M_Z^4 - 12 M_t^4 \right], \\
    y_{t}^2(\mu) &= 2 \sqrt{2} G_F M_t^2 \left[ 1 + \frac{1}{(4\pi)^2} \left( 8 g_s^2 (L_T - \frac{4}{3}) - \sqrt{2} G_\mu M_t^2 ( 9 L_T - 11)\right) \right], \\
    y_{b,\tau}^2 &= 2 \sqrt{2} G_F M_{b,\tau}^2,
\end{align}
where $L_i \equiv \ln{(M_i^2/\mu^2)}$ and masses in capital letters denote pole masses. For the electroweak gauge couplings at Born level,
\begin{align}
    g_Y^2 &= 4 \sqrt{2} G_F (M_Z^2 - M_W^2), \\
    g_2^2 &= 4 \sqrt{2} G_F M_W^2 .
\end{align}

\subsection{Axion-Higgs model}
In the Axion-Higgs model, relations between the $\overline{\text{MS}}$ parameters and the physical parameters are similar to the Standard Model, just that we have to take the mass-mixing into account. As usual, we therefore perform a linear field redefinition via a rotation in the basis of the two scalar fields to diagonalise the quadratic terms in $h$ and $S$. To this end, we compute the \textit{mass matrix} $\mathbf{M}$,
\begin{equation}\label{eq:definition-mass-matrix}
    \mathbf{M}_{ij}(p^2) \equiv \partial_i \partial_j V_\text{eff} + \Delta \Pi_{ij}(p^2).
\end{equation}
with $i,j \in \{H,S\}$ and $\Delta\Pi(p^2)=\Pi(p^2)-\Pi(0)$. The pole masses $M_h$ and $M_S$ are then equal to the eigenvalues of the mass matrix evaluated at $p^2 = M_h^2$ resp.\@ $p^2 = M_S^2$.  Further, we define the mixing angle $\theta$ via
\begin{equation}
    \operatorname{diag}{(M_h^2, M_S^2)} = R(\theta) \cdot \mathbf{M} \cdot R^{-1}(\theta),
\end{equation}
where $R(\theta)$ is the 2D rotation matrix. At one-loop, and ignoring the subdominant corrections due to the 2-point functions,
\begin{equation}
    \operatorname{diag}{(M_h^2,M_S^2)} = R(\theta) \cdot HV_\text{eff}(H,S) \cdot R^{-1}(\theta),
\end{equation}
with the Hessian matrix of the effective potential $HV_\text{eff}(H,S)$. Together with the minimum conditions,
\begin{equation}
    \left.\pdv{V_\text{eff}}{H}\right\rvert_{\substack{H = v \\ S = 0 }} = 0, \quad \left.\pdv{V_\text{eff}}{S}\right\rvert_{\substack{H = v \\ S = 0 }} = 0,
\end{equation}
we can invert these relations for the $\overline{\mbox{MS}}$ parameters of this theory in terms of physical quantities and the mixing angle.\footnote{The mixing angle $\theta$ is not imminently physical but we use it as an input parameter for our analysis of the parameter space.} Note that the second condition is already implicit in our definition of the potential resp.\@ Lagrangian. In particular, from these relations follows that
\begin{equation}
    \sin^2 {\theta} = \frac{(M_h^2 - M_{HH})^2}{(M_h^2 - M_{HH})^2 + (M_{HS})^2},
\end{equation}
where $M_h^2$ is the squared Higgs pole mass, and $M_{HH}, M_{HS}$ are elements of the mass matrix, see Eq.~\eqref{eq:definition-mass-matrix}. At NLO for the top and Higgs parameters, and LO otherwise, we then obtain,
\begin{align}\label{eq:Axion-Higgs-parameter-relations}
    \lambda(\mu) &= \frac{G_F}{\sqrt{2}} \left(M_h^2 c_\theta^2 + M_S^2 s_\theta^2 \right) - \frac{2 G_F^2}{(4\pi)^2} \left[ 6 M_W^4 \left(L_W + \frac{2}{3}\right) + 3 M_Z^4 \left(L_Z + \frac{2}{3}\right) - 12 M_t^4 L_T \right], \\
    m_H^2 (\mu) &= \frac{1}{2} \left(M_h^2 c_\theta^2 + M_S^2 s_\theta^2 \right) - \frac{\sqrt{2} G_F}{(4 \pi)^2} \left[ 6 M_W^4 + 3 M_Z^4 - 12 M_t^4 \right], \\
    A &= (\sqrt{2} G_F)^{1/2} (M_h^2 - M_S^2) s_{\theta} c_{\theta}, \\
    m_S^2 &= M_S^2 c_\theta^2 + M_h^2 s_{\theta}^2, \\
    y_{t}^2(\mu) &= 2 \sqrt{2} G_F M_t^2 \left[ 1 + \frac{1}{(4\pi)^2} \left( 8 g_s^2 (L_T - \frac{4}{3}) - \sqrt{2} G_\mu M_t^2 ( 9 L_T - 11)\right) \right], \\
    y_{b,\tau}^2 &= 2 \sqrt{2} G_F M_{b,\tau}^2,
\end{align}
where $s_\theta \equiv \sin{\theta}, c_\theta \equiv \cos{\theta}$, and for the electroweak gauge couplings as before,
\begin{align}
    g_Y^2 &= 4 \sqrt{2} G_F (M_Z^2 - M_W^2), \\
    g_2^2 &= 4 \sqrt{2} G_F M_W^2 ,
\end{align}
at leading order.

\section{Numerical Values in the Standard Model}
We use the following values for the physical quantities \cite{buttazzo2013investigatingnearcriticalityhiggs,navas2024reviewparticlephysics}, and the strong coupling constant $\alpha_S$ from the PDG 2023 average \cite{buttazzo2013investigatingnearcriticalityhiggs,navas2024reviewparticlephysics},
\begin{equation}\label{eq:PDG-values}
\begin{split}
    M_t &= \SI{173.1(7)}{\GeV},\\
    M_b &= \SI{4.183(7)}{\GeV},\\
    M_\tau &= \SI{1.77693(9)}{\GeV},\\
    M_h &= \SI{125.20(11)}{\GeV},\\
    M_Z &= \SI{91.1880(20)}{\GeV},\\
    M_W &= \SI{80.3692(133)}{\GeV},\\
    G_F &= \SI{1.1663788(6)e-5}{\per\square\GeV}, \\
    \alpha_S(M_t) &= \num{0.1080(8)}
\end{split}
\end{equation}
For a more recent discussion on the top mass determination, see e.g.~Ref.~\cite{navas2024reviewparticlephysics}.
The dominant sources of uncertainty for the determination of the Higgs quartic coupling and its running are the top quark mass and Higgs mass, and the uncertainty in the strong coupling constant. We neglect all other uncertainties in the following as we are working with the leading NLO threshold corrections only.
With the one-loop threshold corrections, using Eq.~\eqref{eq:SM-threshold-corrections} and following, we find at the top mass scale $M_t$,
\begin{equation}\label{eq:couplings-at-top-scale}
\begin{split}
    \lambda(M_t) &= \num{0.12987(24)} ,\\
    g_Y(M_t) &= \num{0.349946} ,\\
    g_2(M_t) &= \num{0.652825} ,\\
    g_3(M_t) &= \num{1.1650(43)} ,\\
    y_t(M_t) &= \num{0.9373(48)} ,\\
    y_b(M_t) &= \num{0.024026} ,\\
    y_\tau(M_t) &= \num{0.010206} ,
\end{split}
\end{equation}
The uncertainties are obtained from varying the input parameters (masses) by one standard deviation. The above uncertainties are the experimental uncertainties only, so do not account for the theoretical uncertainties from the electroweak threshold corrections at full NLO.
We compute the renormalisation group $\beta$-functions for the Standard Model and the Axion-Higgs model at 3-loop using the Mathematica package \texttt{RGBeta} \cite{thomsen2021rgbetamathematicapackage}. We take into account the leading and sub-leading contributions from the top, bottom and $\tau$ Yukawa coupling, and neglect all other Yukawa couplings.
Evolving the top mass scale $M_t$ to the Planck scale $M_\text{Pl}$, we obtain the couplings at the Planck scale from the central values in Eq.~\eqref{eq:couplings-at-top-scale}:
\begin{equation}\label{eq:couplings-at-Planck-scale}
\begin{split}
    \lambda(M_\text{Pl}) &= \num{-0.0102(80)} ,\\
    g_Y(M_\text{Pl}) &= \num{0.45751(1)} ,\\
    g_2(M_\text{Pl}) &= \num{0.50783(1)} ,\\
    g_3(M_\text{Pl}) &= \num{0.48718(31)} ,\\
    y_t(M_\text{Pl}) &= \num{0.3842(67)} ,\\
    y_b(M_\text{Pl}) &= \num{0.00793(5)} ,\\
    y_\tau(M_\text{Pl}) &= \num{0.00947(5)} ,
\end{split}
\end{equation}
where the uncertainties are obtained from varying the couplings at the top mass scale by one standard deviation.
For all other Standard Model parameters relevant to this work we used those quoted by the Particle Data Group \cite{navas2024reviewparticlephysics}.

\newpage
\bibliography{main}
\bibliographystyle{JHEP}

\end{document}